\documentstyle[preprint,revtex,eqsecnum]{aps}
\begin{document}
\draft
\preprint{IP/BBSR/92-76}
\begin{title}
\bf Gluon Condensates, Chiral Symmetry Breaking
\end{title}
\begin{title}
\bf And Pion Wave-function
\end{title}
\author{A. Mishra}
\begin{instit} {Physics Department, Utkal University,
Bhubaneswar-751 004, India.}
\end{instit}
\author {S.P. Misra}
\begin{instit}
Institute of Physics, Bhubaneswar-751 005, India.
\end{instit}
\begin{abstract}
We consider here chiral symmetry breaking in quantum chromodynamics
arising from gluon condensates in vacuum.
Through coherent states of gluons simulating a mean field
type of approximation, we show that the off-shell gluon condensates
of vacuum generate a mass-like contribution for the
quarks, giving rise to chiral symmetry breaking. We next note that
spontaneous breaking of global chiral symmetry links the four component
quark field operator to the pion wave function. This in turn yields
many hadronic properties in the light quark sector in agreement with
experiments, leading to the conclusion that low energy hadron properties
are primarily driven by the vacuum structure of quantum chromodynamics.
\end{abstract}
\newpage

\section {\bf Introduction}
\bibcite {ltg}{1}
\bibcite {creutz}{2}
\bibcite{svz}{3}
It is now accepted that quantum chromodynamics (QCD) is the correct theory
for strong interaction physics of quarks and of hadrons.
However, this being a nonabelian gauge theory of strong interactions,
at present no reliable method other than lattice gauge theory
\cite{ltg} is known for the solution of such problems \cite{creutz}.
Although this is a beautiful and
powerful method, the solutions finally need
better computer capabilities compared to what is available now.
It is thus desirable
to consider alternative dynamical schemes which do not involve
a discretisation of space-time. One such scheme with a
limited objective has been the QCD sum rules \cite{svz} through
duality principle. It optimises the use of
a nonperturbative vacuum structure with perturbative QCD calculations
 \cite{svz},
and reproduces nice experimental results. However, in a purely nonperturbative
situation, the method is expected not to be applicable.

\bibcite {spm87a}{4}
\bibcite {spmbg}{5}
\bibcite {am91}{6}
\bibcite {spmtlk}{7}
\bibcite {hmam}{8}
\bibcite {higgs}{9}
\bibcite {nmtr}{10}
\bibcite {dtrn}{11}
We have proposed earlier an alternative scheme with a variational method
which is
nonperturbative \cite{spm87a,spmbg}, and uses off-shell gluons for the
existence of a nontrivial vacuum structure \cite{am91}.
Calculations are done in Coulomb gauge. The inputs consist of the
QCD Lagrangian, and
QCD vacuum contains gluon condensates arising through a
variational principle \cite{am91}. With a similar
technique we have also considered chiral symmetry breaking in Nambu
Jona Lasinio model, as well as for some
quark antiquark phenomenological potentials used for hadron spectroscopy
\cite{spmtlk,hmam}. In the present paper we consider the same as
arising from gluon condensates along with some low energy hadronic
phenomena.

The paper is organised as follows. In section {\bf 2} we recapitulate
\cite{am91} the nonperturbative gluon condensates in Coulomb gauge
giving rise to a nontrivial vacuum structure and define notations.
We then conjecture that through minimal coupling the quarks have a nonzero
mass-like contribution, and obtain an approximate value for the same.
In section {\bf 3} we derive the pion state as a quark antiquark pair
through a constructional use of Goldstone theorem.
In section {\bf 4} we proceed to use this relationship to obtain some
low energy hadronic properties in the light quark sector.
In section {\bf 5} we discuss the results.

The method considered here
is a non-perturbative one as we use only equal time quantum algebra
but is limited by the choice of ansatz functions in the calculations.
The techniques have been applied earlier to solvable cases to examine
its reliability \cite{spmbg} as well as to physically more relevant ground
state structures in high energy physics \cite{am91,higgs} and nuclear physics
\cite{nmtr,dtrn}. Starting from the QCD Lagrangian
we examine here the more complex problem of chiral symmetry breaking
and its relationship to low energy hadronic properties.

\section{\bf gluon condensates and chiral symmetry breaking}
In the present section we shall recapitulate the nontrivial vacuum
structure in quantum chromodynamics through Bogoliubov transformation
as considered earlier \cite{am91} to set the notations, and derive some
results.

The Lagrangian density including the quark fields is given as
\begin{equation}
{\cal L} ={\cal L} _{gauge}+{\cal L} _{matter}+{\cal L} _{int}.\end{equation}
Here,
\begin{equation}
{\cal L} _{gauge}=-{1\over 2}G^{a\mu\nu}(\partial_{\mu}{W^{a}}_{\nu}
-\partial_{\nu}{W^{a}}_{\mu}+gf^{abc}{W^{b}}_{\mu}{W^{c}_{\nu})}
+{1\over 4}{G^{a}}_{\mu\nu}{G^{a\mu\nu}}.\end{equation}
\noindent where ${W^{a}}_{\mu}$ are the SU(3) colour gauge fields.
Also the Lagrangian densities for the quark fields and interactions
with gluons are given as
\begin{equation}
{\cal L} _{matter}={\bar\psi(i\gamma^{\mu}\partial_{\mu}-m)\psi} ,
\end{equation}
and
\begin{equation}
{\cal L} _{int}=g \bar \psi \gamma ^\mu t_a{W^a}_ \mu \psi .\end{equation}
\bibcite{schwinger}{12}
\bibcite{schutte}{13}
\bibcite{gft}{14}
\bibcite{qcd}{15}
\bibcite{gribov}{16}

In order to do the quantisation in Coulomb gauge, we write the
electric field, ${G^{a}}_{0i}$ in terms of
the transverse and longitudinal parts as \cite{am91}
\begin{equation}
{G^a}_{0i}=
{^TG^a}_{0i}+{\partial_{i}{f}^{a}},\end{equation}
\noindent where the form of ${f}^{a}$ is to be determined.
In the Coulomb gauge the
subsidiary condition and the equal time algebra for the gauge
fields are given as \cite{schwinger}
\begin{equation}
{\partial_{i}}{W^a}_i=0 \end{equation}
and
\begin{equation}
\left [{W^{a}}_{i}({\vec x},t),^{T}{G^{b}}_{0j}
({\vec y},t)\right ]=i{\delta}^{ab}({\delta}_{ij}
-{{\partial_{i}\partial_{j}}\over {\partial^2}})\delta
({\vec x}-{\vec y}).\end{equation}
\noindent We take the field
expansions for ${W^{a}}_{i}$ and $^{T}{G^{a}}_{0i}$
at time t=0 as \cite{schutte}
\begin{mathletters}
\begin{equation}
{W^{a}}_{i}(\vec x)={(2\pi)^{-3/2}}\int
{d\vec k\over \sqrt{2\omega(\vec k)}}({a^a}_{i}(\vec k) +
{{a^a}_{i}(-\vec k)}^{\dagger})\exp({i\vec k\cdot\vec x})\end{equation}
 \noindent and
 \begin{equation}
 {^{T}G^{a}}_{0i}(\vec x)={(2\pi)^{-3/2}} i \int
{d\vec k}{\sqrt{\omega(\vec k)\over 2}}(-{a^a}_{i}(\vec k) +
{{a^a}_{i}(-\vec k)}^{\dagger})\exp({i\vec k\cdot\vec x}).\end{equation}
\end{mathletters}
\noindent From equation (2.7) these give the commutation
relations for ${a^a}_{i}$ and ${{a^b}_{j}}^\dagger$ as
\begin{equation}
\left[ {a^a}_{i}(\vec k),{{a^b}_{j}(\vec k^{'})}^\dagger\right]=
\delta^{ab}\Delta_{ij}(\vec k)\delta({\vec k}-{\vec k^{'}}),\end{equation}
where, $\omega$(k) is arbitrary, and,
 \begin{equation}
 \Delta_{ij}(\vec k)={\delta_{ij}}-{k_{i}k_{j}\over k^2}.
\end{equation}
\noindent In Coulomb gauge, the expression for the Hamiltonian
density, ${\cal T}^{00}$ from equation (2.1) is given as \cite{gft}
\begin{eqnarray}
{\cal T}^{00}&=&:{1\over 2}{^{T}{G^a}_{0i}}{^{T}{G^a}_{0i}}+
{1\over 2}{W^a}_{i}(-\vec \bigtriangledown^2){W^a}_{i}+
g \; f^{abc}{W^b}_i{W^c}_j \partial _i{W^a}_j \nonumber\\
&+&{{g^2}\over 4}f^{abc}f^{aef}{W^b}_{i}{W^c}_{j}{W^e}_{i}{W^ f}_{j}
+ {1\over 2}(\partial_{i}f^{a})(\partial_{i}f^{a})\nonumber\\
&+&\bar \psi (-i \gamma ^i \partial _i-m)\psi
-g \bar \psi \gamma ^i t_a{W^a}_i \psi :,\end{eqnarray}
\noindent where : : denotes the normal ordering with respect to
the perturbative vacuum, say $\mid vac>$, defined through
${a^a}_{i}(\vec k)\mid vac>=0$. The term ${1 \over 2}(\partial _if^a)
(\partial _if^a)$ automatically includes interactions for both
time-like and longitudinal gluons, here through the auxiliary field
description, and ${\cal T}^{00}$ is calculated after the elimination
of the same \cite{am91}.

In Ref [3] the possibility of a nonperturbative vacuum having a lower energy
 was discussed as from the above equation.
 Such a vacuum contained gluon condensates. We thus considered earlier
 a trial state, $\mid vac^{'}>$ with such condensates
over the perturbative vacuum, $\mid vac>$ given as \cite{spmbg,am91}
\begin{equation}
\mid vac ^{'}>
=U\mid vac>,\end{equation}
\noindent where \begin{equation}
U=\exp({B^\dagger}-B),
\end{equation}
\noindent with
\begin{equation}
{B^\dagger}={1\over 2}
\int {f(\vec k){{{a^a}_{i}(\vec k)}^\dagger}
{{{a^a}_{i}(-\vec k)}^{\dagger}}d\vec k}.\end{equation}
$B ^\dagger $ contains gluon pair creation operators with an ansatz
function $f(\vec k)$ to be determined later.
The operators, say,
${b^a}_{i}(\vec k)$, which annihilate $\mid vac^{'}>$ are
given as
\begin{equation}
{b^a}_{i}(\vec k)=U{a^a}_{i}(\vec k){U}^{-1}.
\end{equation}
We can explicitly evaluate from the equations above the operators
${b^a}_{i}(\vec k)$ in terms of ${a^a}_{i}(\vec k)$. In fact,
we then obtain the Bogoliubov transformations \cite{am91}
\begin{equation}
\pmatrix {{b^a}_{i}(\vec k) \cr {{b^a}_{i}(-\vec k)}^\dagger}
=\pmatrix {\cosh f(\vec k) & -\sinh f(\vec k) \cr
-\sinh f(\vec k) & \cosh f(\vec k)}\pmatrix {{a^a}_{i}(\vec k)  \cr
{{a^a}_{i}(-\vec k)}^\dagger},\end{equation}
where we have assumed that the function $f(\vec k)$ is even and real.
Using equations (2.9) and (2.16), we obtain the same
commutation relation for the operators ${b^a}_{i}$
and ${{b^b}_{j}}^\dagger$ given as
\begin{equation}
\left[ {b^a}_{i}(\vec k),
{{b^b}_{j}(\vec k^{'})}^\dagger \right]
=\delta^{ab}\Delta_{ij}(\vec k)\delta({\vec k}-{\vec k^{'}}),
\end{equation}
which merely reflects that the Bogoliubov
transformation (2.16) is a canonical transformation.
In order to consider the stability of $\mid vac>$ with respect to the
transformation of equation (2.12), the expectation value of
${\cal T}^{00}$ with respect to $\mid vac'>$ was evaluated and was
then minimised over $f(\vec k)$.

It was noticed \cite{am91} that under certain conditions the
perturbative vacuum becomes unstable, where it was also seen that the general
solution of $f(\vec k)$ through minimisation is impossible to obtain
analytically. We had taken an ansatz function $f(\vec k)$ of
equation (2.14) for $\mid vac'>$ description as given by \cite{am91}
\begin{equation}
sinh(f(\kappa ))=A \; e^{-B \kappa ^2/2}\end{equation}
and then had determined the parameter $A$ through a minimisation of
vacuum energy density. The dimensional parameter $B$ was determined
through the evaluation of the SVZ parameter \cite{am91}.
We may note from equations (2.16) and (2.18) that
\begin{equation}
<vac'|{a^a}_i(\vec k)^\dagger {a^a}_i(\vec k')|vac'>
=16.A^2 exp(-B \kappa ^2)\delta (\vec k-\vec k'),\end{equation}
so that the ansatz of equation (2.18) merely implies a Gaussian
distribution for the perturbative gluons in $|vac'>$.
The minimisation was done for pure gluon fields in Ref. \cite{am91}
with $\alpha_s=g^2/(4\pi)$ of the Lagrangian as the input coupling strength.
For example, we then have for input values of
$\alpha _s=0.5,\; 0.6,\; 0.8,\; 1.0,\;$ the identification for the
output parameters as $A_{min}=.178,\; .306,\; .557,\; .658,\;$
$ B^{1 \over 2}=.40,\; .48,\; .62,\; .69\; $
 in fermis, and $\; m=232,\; 272,\; 325,\; 345\;$ in MeV respectively, where
 $\omega (\kappa )=\sqrt {\kappa ^2+m^2}$. $``$m" here is a mass-like
 parameter for $\omega (\kappa)$ of equations (2.8) which was determined
 through a self-consistent calculation.
It is {\it not} associated with energy momentum four vector of a
$``$free'' gluon, which anyhow does not exist. We may note that the
results of Ref. \cite{am91} are very similar to some of the earlier
results \cite{qcd}, and that the method of solution for the auxiliary
equation avoided Gribov ambiguity \cite{gribov}.

Here we shall be using the final output,
the stable vacuum $\mid vac'>$ as the physical vacuum.
We shall assume that the vacuum structure in QCD in presence of quarks
is mainly driven by gluon condensates, and study some consequences
of the same.

We now proceed to consider the quark fields in the presence of the
$``$dressing" of vacuum with off-mass shell gluon quanta
as considered above. We note that here with minimal
coupling
\begin{equation}
\vec k^2\rightarrow\vec k^2+<vac'|g^2t^at^b{W^a}_i{W^b}_i|vac'>.
\end{equation}
With the mean field approximation for the gluon fields as above,
we then see that the mass of the quarks $m_Q$ is now given as
\begin{equation}
m_Q^2\approx m^2+g^2 t_at_a f_{ii}(\vec 0),
\end{equation}
where we have substituted \cite{am91}
\begin{eqnarray}
<vac'\mid {W^a}_i(\vec x ){W^b}_j(\vec y )\mid vac'>
&=&\frac{\delta ^{ab}}{(2  \pi )^3}\int d\vec k e^{i\vec k.(\vec x-
\vec y)}\; {F_{+}(\vec  k)\over \omega (k)}\;
\Delta _{ij}(\vec k)\nonumber\\
&\equiv &\delta^{ab} \; f_{ij}(\vec x -\vec y ).
\end{eqnarray}
In the above \cite{am91}
\begin{equation}
F_{+}(\vec k)=\biggl({\sinh 2f(k)\over 2} + {\sinh}^{2}f(k)\biggr).
\end{equation}
With $t_at_a=4/3$, we then obtain that
\begin{equation}
m_Q^2=m^2+\frac{16\pi\alpha_s}{3}f_{ii}(\vec 0)
\end{equation}
where summation over the index $i$ is understood.
We thus note that even when $m=0$ there is a finite mass of the quark,
which implies breaking of chiral symmetry. We then obtain through
numerical evaluation \cite{am91} of $f_{ii}(\vec 0)$ that
$m_Q^2=.0358$ GeV$^2$ for $\alpha_s=0.5$, so that we then have
\begin{equation}
m_Q=189 {\rm MeV}.
\end{equation}

We may however note that the above result is deceptive.
It is incomplete since we have made the mean field approximation with
minimal coupling to generate a mass-like contribution, and do not know
how it operates regarding details of dynamics. We use this only to
illustrate that gluon condensates break chiral symmetry.
We shall include this effect in the expansion of the quark field
operators.

\bibcite{spm78}{17}
\bibcite{lopr}{18}
\bibcite{sp86}{19}
We then write the quark fields as
\begin{equation}
\psi (\vec x )=\frac{1}{(2\pi)^{3/2}}\int \left[U(\vec k)q_I(\vec k )
+V(-\vec k)\tilde q_I(-\vec k )\right]e^{i\vec k\cdot \vec x}d \vec k,
\end{equation}
where \cite{spm78,lopr}
\begin{mathletters}
\begin{equation}
U(\vec k )=\pmatrix {cos\frac{\chi(\vec k)}{2} \cr
\vec \sigma \cdot \hat k sin\frac{\chi(\vec k)}{2}\cr},
\end{equation}
and
\begin{equation}
V(-\vec k )=\pmatrix {-\vec \sigma \cdot \hat k
sin\frac{\chi(\vec k)}{2} \cr cos\frac{\chi(\vec k)}{2}\cr}.
\end{equation}\end{mathletters}
The above form is so taken that it satisfies the equal time algebra
\cite{spm78,lopr}.
In the above, for free quark fields of mass $m_Q$, we have \cite{spm78}
\begin{equation}
cos\frac{\chi(\vec k)}{2}=\Big({p^0+m_Q \over 2p^0}\Big)^{1 \over 2},
\qquad
sin\frac{\chi(\vec k)}{2}=\Big({p^0-m_Q \over 2p^0}\Big)^{1/2},
\end{equation}
with $p^0=\big(m^2_Q+\vec k^2 \big)^{1 \over 2}$.
We may further note that {\it when chiral symmetry is unbroken}
$\chi(\vec k)=\pi/2$. However, since the quarks are not free, and,
chiral symmetry is broken, $\chi(\vec k)$ shall be different from the
above.

We would like to associate the quark field operators as in equation
(2.26) explicitly with the presence of quark condensates generating
chiral symmetry breaking. Parallel to the earlier operators for
gluon condensates, let us consider the quark antiquark pair creation
operator, with $h(\vec k)$ as a trial function,
\begin{equation}
B_Q ^\dagger =\int h(\vec k)q_I(\vec k)^\dagger \vec\sigma\cdot\hat k
\tilde q_I(-\vec k)d\vec k.
\end{equation}
Let us now consider the unitary transformation
\begin{equation}
U_Q=e^{B_Q ^\dagger -B_Q},
\end{equation}
which will operate on chiral vacuum. Also, let us write the corresponding
quark field operator of equation (2.26) in momentum space as
\begin{equation}
\tilde \psi_0(\vec k)=\left(\begin{array}{c}cos\frac{\chi_0}{2}q_I(\vec k)
-\vec \sigma\cdot\hat  k sin\frac{\chi_0}{2}\tilde q_I(-\vec k)\\
\vec \sigma\cdot\hat  k sin\frac{\chi_0}{2} q_I(\vec k)+
cos\frac{\chi_0}{2}\tilde q_I(-\vec k)\end{array} \right).
\end{equation}
When chiral symmetry is there, $\chi_0=\pi/2$.
We may now note that the unitary transformation $U_Q$ is equivalent
to the Bogoliubov transformation given as
\begin{equation}
U_Q ^\dagger \left(\begin{array}{c}q_I(\vec k)\\
\tilde q_I(-\vec k)\end{array} \right)U_Q=
\left(\begin{array}{cc}cos(h(\vec k))& \vec \sigma\cdot\hat k sin(h(\vec k))\\
-\vec \sigma\cdot\hat k sin(h(\vec k))& cos(h(\vec k))\end{array} \right)
\left(\begin{array}{c}q_I(\vec k)\\
\tilde q_I(-\vec k)\end{array} \right).
\end{equation}
Now, with $\tilde\psi(\vec k)=U_Q ^\dagger
\tilde\psi_0(\vec k) U_Q$, we generate equation (2.26) when we identify
that
\begin{equation}
\frac{\chi(\vec k)}{2}=\frac{\chi_0}{2}-h(\vec k).
\end{equation}
We thus note that the form of equation (2.26) with (2.27) corresponding
to chiral symmetry breaking can be interpreted as a destabilisation
of chiral symmetric vacuum through equation (2.30) when $\chi_0=\pi/2$.
{\it Hence taking the quark field operators with chiral
symmetry breaking or taking a condensate over chiral vacuum become
equivalent}. The quark field operators after chiral symmetry breaking
get related to the vacuum structure.

We may also note that
\begin{equation}
<\bar\psi(\vec x)\psi(\vec y)>=
<vac|U_Q ^\dagger \bar \psi_0(\vec x)\psi_0(\vec y)U_Q|vac>
=-\frac{12}{(2\pi)^3}\int cos\chi(\vec k)e^{i\vec k\cdot(\vec x -
\vec y)}d \vec k,
\end{equation}
where, the factor 12 arises from colour, flavour and spin degress of
freedom. We may substitute $cos\chi(\vec k)=
sin(2h(\vec k))$, describing the above function explicitly in terms of
the quark antiquark correlations. We also have
\begin{equation}
<\bar\psi(\vec x)\psi(\vec x)>=-\mu^3
\end{equation}
with $\mu$ determined in terms of the condensate function or
$cos\chi(\vec k)$. The condensate function $h(\vec k)$ vanishes when
$cos\chi(\vec k)=0$, as should be the case for chiral vacuum.

The form of $\chi(\vec k)$ will, generally speaking,
depend on interactions \cite{spmtlk,lopr},
and, can be obtained through an extremisation \cite{hmam,sp86}.
Also, the function $\chi(\vec k)$ will get related to
the pion wave function using Goldstone
theorem as shown below. This shall include vacuum structure as a
{\it post facto} information linked to phenomenology without
an extremisation of energy density containing highly nonlinear
expressions \cite{am91}.

\section {\bf Chiral symmetry breaking and pion wave function}
We shall here first consider only two quark flavours and, the usual
three colours.
When chiral symmetry remains good, we note that \cite{spmtlk,hmam}
\begin{equation} {Q_5}^i \mid vac >=0 \end{equation}
where ${Q_5}^i$ is the chiral charge operator given as
\begin{equation}
{Q_5}^i =\int \psi(\vec x)^{\dagger}\frac{\tau^i}{2}\gamma ^5\psi(\vec x)
d \vec x.
\end{equation}
In this case quarks are massless and we have
\begin{equation}
cos\frac{\chi(\vec k)}{2}= sin\frac{\chi(\vec k)}{2}=\frac{1}{\sqrt 2}.
\end{equation}
For symmetry broken case however \cite{spmtlk,hmam}
\begin{equation}
{Q_5}^i\mid vac'>\not = 0.
\end{equation}
This will correspond to the pion state. To show this we first note that
\begin{equation} \left[{Q_5}^i, H\right]=0.
\end{equation}
Clearly, for the symmetry breaking phase, $|vac'>$ is an approximate
eigenstate of the total Hamiltonian $H$ with
${\cal E} V$ as the approximate eigenvalue, $V$ being the total volume and
${\cal E}$ the energy density \cite{hmam}.
With $H_{eff}=H-{\cal E} V$, we then obtain from equation (3.5) that
\begin{equation}
H_{eff}{Q_5}^i|vac'>=0
\end{equation}
i.e. the state ${Q_5}^i\mid vac'>$ with zero momentum
has also zero energy, which corresponds to the massless pion
\cite{spmtlk,hmam}.
Explicitly using equations (2.26) and (2.27), and
with $q_I$ now  as two component isospin doublet corresponding to
(u,d) quarks above, we then obtain the pion states of zero momentum as
\begin{equation}
\mid \pi^i(\vec 0)>=N_{\pi}\cdot\frac{1}{\sqrt6}\int q_I(\vec k)^\dagger
\tau^i {\tilde q_I}(-\vec k)cos(\chi(\vec k))
d\vec k \mid vac'>,
\end{equation}
where, $N_\pi$ is a normalisation constant.
The wave function for pion $\tilde u_{\pi}(\vec k)$ thus is given as
proportional to $cos\chi(\vec k)$.
The colour, isospin and spin indices of $q^\dagger$ and $\tilde q$ for
quarks and antiquarks have been suppressed. Further, with
\begin{equation}
<\pi^i(\vec 0)\mid \pi^j(\vec p)>=\delta^{ij}\delta(\vec p),
\end{equation}
the normalisation constant $N_\pi$  is given through \cite{spm78}
\begin{equation}
{N_{\pi}}^2 \int cos^2(\chi(\vec k)) d\vec k =1.
\end{equation}
In the notations of Ref. \cite{spm78} thus the pion wave function
$\tilde u_{\pi}(\vec k)$ is given as
\begin{equation}
\tilde u_{\pi}(\vec k)=N_{\pi}cos(\chi(\vec k)).
\end{equation}

Clearly the state as in equation (3.7) as the Goldstone mode
will be
accurate to the extent we determine the vacuum structure sufficiently
accurately through
variational or any other method so that $|vac'>$ is
 an eigenstate of the total Hamiltonian. The above results yield pion
wave function from the vacuum structure for any
example of chiral symmetry breaking, and is a new feature of looking at
phase transition through vacuum realignment \cite{spmtlk}.

We had earlier considered low energy hadronic properties \cite{spm78}
with an assumed form of interacting quark field operators. This is
equivalent to a choice of $cos(\chi(\vec k))$. In addition, we had
taken Gaussian wave functions for baryons and mesons, and then discussed
the phenomenology. We shall here examine the consistency of the present
picture with low energy hadron phenomenology where pion wave function
and quark field operators get related as in equations (2.27) and (3.10).

The notations of Ref. \cite{spm78}, which we shall use now, correspond to
\begin{equation}
f_q(\vec k)=cos\frac{\chi(\vec k)}{2};\qquad
|\vec k|g_q(\vec k)=sin\frac{\chi(\vec k)}{2}.
\end{equation}
We may note that as in equation (2.26) with (2.27), we are retaining
the fully relativistic four-component quark field operator. {\it The
phenomenological assumption here will consist of explicitly taking a
specific expression for} $cos\chi(\vec k)$. We note that $cos\chi(\vec k)$
as in equation (2.34) is the correlation function for quark pairs in
vacuum. Hence, parallel to the choice of gluon correlation functions
in equation (2.19) as a Gaussian \cite{am91}, we shall take here the
ansatz
\begin{equation}
cos\chi(\vec k)=exp\left(-\frac{R_{\pi}^2}{2}\vec k^2\right).
\end{equation}
It also implies through equation (3.10) that the pion wave
function is a Gaussian. From equation (3.9), here we have
\begin{equation}
N_{\pi}=\frac{R_{\pi}^{3/2}}{\pi^{3/4}} =0.424\times R_{\pi}^{3/2}.
\end{equation}
 From equations (2.34) and (2.35) we also obtain that
\begin{equation}
\mu=\frac{(12)^{1/3}}{\sqrt{2\pi}}R_{\pi}^{-1}=0.913\times R_{\pi}^{-1}.
\end{equation}

We would now like to see what may be the nature of dispersion curves
for quarks in vacuum.
We note that for free quark fields $sin\chi(\vec k)=\kappa/\epsilon(
\vec k)$. We may use this to obtain that
\begin{equation}
\epsilon(\vec k)=\frac{\kappa}{\sqrt{1-cos^2\chi(\vec k)}},
\end{equation}
{\it giving the dispersion curve for quarks in vacuum as a medium}.
Thus for equation (3.12),
\begin{mathletters}
\begin{equation}
\epsilon(\vec k)=\frac{\kappa}{\sqrt{1-e^{R_{\pi}^2\vec k^2}}}.
\end{equation}
Hence, for small $\kappa^2$,
\begin{equation}
\epsilon(\vec k)=\frac{1}{R_{\pi}}+\frac{R_{\pi}}{4}\kappa^2.
\end{equation}
\end{mathletters}
We thus see that $\epsilon(\vec k)$ {\it does not have the form of
a free particle}.
We note that  for large $\kappa$, $\epsilon(\vec k)\approx \kappa$,
which follows for {\it any} choice of $cos\chi(\vec k)$ from
equation (3.9).

We shall need later expansion of $cos\chi(\vec k+\vec p)$ for small
$|\vec p|$ for charge radius and magnetic moment. For this purpose we
use the notations
\begin{equation}
\partial_icos\chi(\vec k)=-\vec k_i\;b_1(\vec k);\qquad
\partial_i\partial_jcos\chi(\vec k)=-\delta_{ij}\;b_1(\vec k)+
\vec k_i\vec k_j b_2(\vec k).
\end{equation}
In the above, we have used rotational invariance. We also note that
\begin{equation}
\vec k^2b_1(\vec k)^2=(\vec \bigtriangledown cos\chi(\vec k))^2,
\end{equation}
and, define $c_2(\vec k)$ such that
\begin{equation}
b_1(\vec k)-\frac{1}{3}\vec k^2b_2(\vec k)\equiv c_2(\vec k)=-\frac{1}{3}
\vec \bigtriangledown ^2cos\chi(\vec k).
\end{equation}
The above equations shall be useful if we wish to change the ansatz
function for $cos\chi(\vec k)$. Clearly, for equation (3.12), we have
\begin{equation}
b_1=R_{\pi}^2cos\chi(\vec k);\qquad b_2=-R_{\pi}^4cos\chi(\vec k).
\end{equation}

\bibcite{spm80a}{20}
\bibcite{spm80b}{21}
\bibcite{arp92}{22}
\bibcite{van}{23}
\section {\bf Low energy hadronic properties}
In the context of chiral symmetry breaking,
the model of Ref. \cite{spm78} now has got modified since
the pion wave function and the quark field operators in the light quark
sector get related. This in many ways enriches
the earlier model \cite{spm78}, as it decreases the number of independent
quantities. The previous model had also been extended to incoherent
processes involving structure function as well as hadronisation
\cite{spm80a,spm80b}. It has also been applied to the problem of strong CP
violation recently \cite{arp92}. We here consider different low energy
hadronic properties parallel to \cite{spm78} using equations (3.11).
\medskip

\subsection{\bf Pion decay constant}
The pion decay constant $c_{\pi}$ as calculated earlier in terms of the
wave function with relativistic correction to Van Royen-Weisskopf
relation \cite{van} is given by \cite{spm78}
\begin{equation}
\left|(1+2g_q^2 \vec \bigtriangledown ^2)u_{\pi}(\vec 0)\right|
=\frac{c_{\pi}(m_{\pi})^{1/2}}{\sqrt 6},
\end{equation}
where the value of the wave function is taken at the space origin, and,
$g_q$ is a differentiation operator. In fact
taking the Fourier transform the left hand side of equation (4.1)
becomes
\[
\frac{1}{(2\pi)^{3/2}}\int (1-2g_q(\vec k)^2\vec k^2)u_{\pi}(\vec k)d \vec k.
\]
 From equation (3.11) however $1-2g_q(\vec k)^2\vec k^2=cos(\chi(\vec k))$.
Hence from equation (3.9) the
normalisation constant $N_{\pi}$ is given through \cite{spm78},
with $c_{\pi}=92$ MeV,
\begin{equation}
\frac{1}{(2\pi)^{3/2}}\cdot\frac{1}{N_{\pi}}
=\frac{c_{\pi}(m_{\pi})^{1/2}}{\sqrt 6}\approx 0.0140 {\rm GeV}^{3/2},
\end{equation}
so that in equation (3.10)
\begin{equation}
N_{\pi}=4.55\;GeV^{-3/2}.
\end{equation}
Hence, from equations (3.13) and (3.14) we may note that, for the choice
of equation (3.12),
\begin{equation}
R_{\pi}^2=\pi\times N_{\pi}^{4/3}=23.69\;{\rm GeV}^{-2};
\qquad \mu=\;187\;{\rm MeV}.
\end{equation}

We may note that in equation (3.16b) if we take $m_Q=1/R_{\pi}$,
then $m_Q$ for the above equation is around 205 MeV, which may be
compared with the earlier value of 189 MeV from the vacuum structure
in QCD due to gluon condensates as in equation (2.25).

Thus the known value of the pion decay constant puts a severe constraint
both on the pion wave function as well as on the form of the quark field
operators.
\medskip

\bibcite{rchpi}{24}
\subsection{\bf Pion charge radius}
We note that the pion form factor is given by
\cite{spm78}
\begin{equation}
<\pi^+(-\vec p)|J^0(0)|\pi^+(\vec p)>=\frac{1}{(2\pi)^3}\frac{m_{\pi}}
{p^0}G^{\pi}_E(t)
\end{equation}
where through direct evaluation \cite{spm78} we obtain that
\begin{equation}
G_E^{\pi}(t)=\int \tilde u_{\pi}(\vec k'_1)^\dagger \tilde u_{\pi}(\vec k_1)
\left(f_q(\vec k_1')f_q(\vec k_1)+\vec k_1'\cdot\vec k_1g_q(\vec k_1')
g_q(\vec k_1)\right)d\vec k_1.
\end{equation}
In the above \cite{spm78},
\begin{equation}
t=-4\vec p^2; \qquad \vec k_1'=\vec k_1-\frac{m_{\pi}}{p^0}\vec p.
\end{equation}
We shall here retain only terms upto $|\vec p|^2$
and use
\begin{equation}
G_E(t)=1+\frac{R_{ch}^2}{6}t\;.
\end{equation}
For the expansion in powers of $t$, let us substitute in equation (4.6)
\begin{equation}
\vec k_1'=\vec k -\frac{1}{2}\vec p; \qquad
\vec k_1=\vec k +\frac{1}{2}\vec p
\end{equation}
corresponding to equation (4.7). Using equations (3.17) and (3.19),
we get that
\begin{equation}
cos\chi(\vec k_1')cos\chi(\vec k_1)=
cos^2\chi(\vec k) -\frac{|\vec p|^2} {4}
\left(cos\chi(\vec k)c_2(\vec k)+\frac{1}{3}\vec k^2b_1(\vec k)^2\right).
\end{equation}
and,
\begin{equation}
cos\chi(\vec k_1')+cos\chi(\vec k_1)=2cos\chi(\vec k)
-\frac{|\vec p|^2} {4} c_2(\vec k).
\end{equation}
We then have in the lowest order in $|\vec p|$,
\begin{equation}
\tilde u_{\pi}(\vec k_1')^\dagger\tilde u_{\pi}(\vec k_1)
=N_{\pi}^2\left[
cos^2\chi(\vec k) -\frac{|\vec p|^2} {4}
\left(cos\chi(\vec k)c_2(\vec k)+\frac{1}{3}\vec k^2b_1(\vec k)^2\right)
\right].
\end{equation}
Also, by equations (3.11) and (4.9),
\begin{equation}
f_Q(\vec k_1')f_Q(\vec k_1)
=\frac{1+cos\chi(\vec k)}{2}
-\frac{|\vec p|^2}{16}\left(c_2(\vec k)+\frac{\vec k^2b_1(\vec k)^2}
{3(1+cos\chi(\vec k))}\right),
\end{equation}
with $\kappa_1'=|\vec k_1'|$ and $\kappa_1=|\vec k_1|$,
\begin{equation}
\kappa_1'g_Q(\vec k_1')\kappa_1g_Q(\vec k_1)
=\frac{1-cos\chi(\vec k)}{2}
-\frac{|\vec p|^2}{16}\left(-c_2(\vec k)+\frac{\vec k^2b_1(\vec k)^2}
{3(1-cos\chi(\vec k))}\right),
\end{equation}
and,
\begin{equation}
\hat k_1'\cdot\hat k_1=1-\frac{\vec p^2}{6\vec k^2}.
\end{equation}
We then easily obtain from equations (4.6), (4.7) and (4.8) that
\begin{equation}
R_{ch}^2=R_1^2+R_2^2
\end{equation}
where, using equations (4.12), (4.13), (4.14) and (4.15)
and simplifying,
\begin{equation}
R_1^2=\frac{N_{\pi}^2}{4}\int (\vec \bigtriangledown cos\chi(\vec k))^2
d\vec k
\end{equation}
is the contribution coming from the wave function alone, and,
\begin{equation}
R_2^2=\frac{N_{\pi}^2}{16}\int cos^2\chi(\vec k)
\left[\frac{\vec k^2b_1(\vec k)^2}{1-cos^2\chi(\vec k)}
+\frac{2(1-cos\chi(\vec k))}{\vec k^2}\right]d\vec k
\end{equation}
is the balance of the contribution.

For the choice of equation (3.12) we then have
\begin{equation}
R_{ch}^2=\frac{3}{8}R_{\pi}^2+I\times R_{\pi}^2,
\end{equation}
where $I=.0159$ as numerically evaluated. This yields that $R_{ch}^2
=9.26$ GeV$^{-2}$, or, $R_{ch}=.605$ fms, which may be compared with
the experimental
value of $R_{ch}=0.66$ fms or, $R_{ch}^2=11.22\;
{\rm GeV}^{-2}$ \cite{rchpi}.
\medskip

\subsection{\bf  $|$g$_A$/g$_V|$}
We next consider some properties associated with proton and neutron.
Let us approximate the proton state with a harmonic oscillator wave
function, so that we take \cite{spm78}
\begin{equation}
\tilde u_p(\vec k_1,\vec k_2,\vec k_3)=\left(\frac{3R_p^4}{\pi^2}\right)
^{3/4}exp\left(-\frac{R_p^2}{6}\sum_{i<j}(\vec k_i-\vec k_j)^2\right).
\end{equation}
With the relativistic corrections, we then
obtain that \cite{spm78}
\begin{equation}
|g_A/g_V|=\frac{5}{3}\cdot\left(\frac{3R_p^2}{2\pi}\right)^{3/2}\int
exp\left(-\frac{3R_p^2}{2}\vec k^2\right)\left(\frac{1}{3}
+\frac{2}{3}cos\chi (\vec k)\right)d \vec k.
\end{equation}
We have obtained here the above expression through a direct
evaluation of
\[<p_{1/2}(\vec 0)|J^+_{5\mu}|n_{1/2}(\vec 0)>\]
retaining the four component quark fields in the current through
equation (2.27), and have substituted from equation (3.11) that
$ f_q^2-\frac{1}{3}g_q^2\vec k^2
\equiv \frac{1}{3}(1+2cos(\chi(\vec k)))$.
Taking $f_q=1$ (and hence $g_q=0$) or, $cos\chi(\vec k)=1$, will give the
result for nonrelativistic quark model. Here the full relativistic
structure of the quark field operator is being
retained. $|g_A/g_V|=1.257$
puts a constraint on the quark field operators and
the approximate size of the nucleons as below.

In fact, we can use the above to obtain the value of $R_p^2$
for the choice of equation (3.12). We note that we then have
\begin{equation}
1.257=\frac{5}{9}\left[1+2\left(\frac{3a}{3a+1}\right)^{3/2}
\right],
\end{equation}
where we have taken $R_p^2=aR_{\pi}^2$. This yields that $R_p^2
=.5707\times R_{\pi}^2$ = 13.52 GeV$^{-2}$.
\medskip

\subsection{\bf Proton charge radius}
In order to make another use of the size of the proton, let us calculate
its charge radius with relativistic corrections.
The charge radius is given through the same identification
as in equation (4.8) with \cite{spm78}
\begin{equation}
<p_r(-\vec p)|J^0(0)|p_s(\vec p)>=\delta_{rs}\frac{1}{(2\pi)^3}\frac{m_p}
{p^0}G^p_E(t).
\end{equation}
In the above, a direct evaluation yields that, upto order $|\vec p|^2$,
\begin{eqnarray}
G^p_E(t)&=&
\left(\frac{3R_p^2}{2\pi}\right)^{3/2}\int
exp\left(-\frac{3R_p^2}{4}(\vec k_1'^2+\vec k_1^2)\right)\nonumber\\
&&\times\left[f_q(\vec k_1')f_q(\vec k_1)+\vec k_1'\cdot\vec k_1g_q(\vec k_1')
g_q(\vec k_1)\right]d\vec k.
\end{eqnarray}
Here, in contrast to equation (4.9), we have \cite{spm78}
\begin{equation}
\vec k_1'=\vec k -\frac{2}{3}\vec p; \qquad
\vec k_1=\vec k +\frac{2}{3}\vec p.
\end{equation}
We then easily see as before that the
proton charge radius is given as
\begin{equation}
R_{ch}^2=R_1^2+R_2^2
\end{equation}
where $R_1^2=R_p^2$ is the nonrelativistic contribution from the proton
wave function as may be
seen with a contribution parallel to that of equation (4.17), and,
\begin{eqnarray}
R_2^2&=&\frac{1}{9}
\left(\frac{3R_p^2}{2\pi}\right)^{3/2}\int
exp\left(-\frac{3R_p^2}{2}\vec k^2\right)\nonumber\\
&\times &
\left[\frac{\vec k^2b_1(\vec k)^2}{1-cos^2\chi(\vec k)}
+\frac{2(1-cos\chi(\vec k))}{\vec k^2}\right]d\vec k
\end{eqnarray}
is the $``$relativistic" contribution.

We have determined $R_p^2$ in the previous subsection. Substituting
the same and using the form of $cos\chi(\vec k)$ as earlier, we obtain
through a numerical evaluation that $R_2^2=3.08$ GeV$^{-2}$, so that
we obtain $R_{ch}^2= 16.6$ GeV$^{-2}$. This gives that
\begin{equation}
R_{ch}=.81\;{\rm fms.,}
\end{equation}
which is similar to the experimental value of the same.
\medskip

\subsection{\bf Proton magnetic moment}
We identify the magnetic moment $\mu_p$ of the proton through the equation
\cite{spm78}
\begin{equation}
\frac{i}{(2\pi)^3}\left[\vec\sigma\times(-2\vec p)\right]_{rs}\mu_p
=<p_r(-\vec p)|J^i(0)|p_s(\vec p)>,
\end{equation}
where on the right hand side above all terms involving $|\vec p|^2$ will
be neglected. Such an evaluation in fact through the notations of
Ref. \cite{spm78} yields that
\begin{equation}
i\left[\vec\sigma\times(-2\vec p)\right]_{rs}\mu_p=
\left(\frac{3R_p^2}{2\pi}\right)^{3/2}\int
exp\left(-\frac{3R_p^2}{2}\vec k^2\right)d\vec k
\bar u^{L^{-1}(p)}_r(\vec k_1')\gamma^i u^{L(p)}_s(\vec k_1).
\end{equation}
In the above, the momenta $\vec k_1$ and $\vec k_1'$ are given by
equatin (4.25) as for
the charge radius, and, $u^{L(p)}(\vec k)=S(L(p))u(\vec k)$
includes spin rotations of the quarks \cite{spm78}. On carrying out
the simplification of the above expression with a straightforward but
complicated algebra, we obtain that
\begin{equation}
\mu_p=\mu_{p1}+\mu_{p2},
\end{equation}
where
\begin{equation}
\mu_{p1}=
\frac{e}{2m}\left(\frac{3R_p^2}{2\pi}\right)^{3/2}\int
exp\left(-\frac{3R_p^2}{2}\vec k^2\right)
\left(\frac{1}{3}+\frac{2}{3}cos\chi(\vec k)\right)d\vec k
\end{equation}
and,
\begin{equation}
\mu_{p2}=
\frac{e}{9}\left(\frac{3R_p^2}{2\pi}\right)^{3/2}\int
exp\left(-\frac{3R_p^2}{2}\vec k^2\right)
\left(\frac{2sin\chi(\vec k)}{\kappa}+\frac{\kappa b_1(\vec k)}
{sin\chi(\vec k)}\right)d \vec k.
\end{equation}
In the above we have used equation (4.13) and (4.14) as earlier.
We may note that $\mu_{p1}$ above came
from Lorentz boosting and in the nonrelativistic limit equals one
nuclear magneton. $\mu_{p2}$ in the nonrelativisic limit
equals (2/3) of quark magnetic moment as in the earlier calculation
\cite{spm78}. We may note that
\begin{equation}
\mu_{p1}=\frac{e}{2m}\times
\frac{3}{5}\frac{g_A}{g_V}.
\end{equation}
With the evaluation of the integral in equation (4.33) using equation
(3.12), we obtain from equation (4.31) that
\begin{equation}
\mu_p=\frac{e}{2m}\times 2.88.
\end{equation}
and, from symmetry or through direct evaluation \cite{spm78},
\begin{equation}
\mu_n=\frac{e}{2m}\times -1.92.
\end{equation}
These results are higher than expected, but still in reasonable agreement
with experiments.
\medskip

\bibcite{chrl}{25}
\bibcite{spm87b}{26}
\section {\bf Discussions}
\medskip

We should note that the present approach to chiral symmetry breaking
through vacuum destabilisation \cite{spmtlk,hmam} is different from
the earlier attempts \cite{chrl}, which mainly use Schwinger Dyson
equations to get the gap equation, and current algebra to
obtain some additional results. We have more conclusions because
of the explicit construction of vacuum state, here even relating the
low energy properties of hadrons to a distribution function for
quarks and antiquarks in vacuum!

Our results here are only linked to
the vacuum structure. There will be additional dressing of the hadrons
by gluons \cite{spm87a,spm87b} which will affect the wave function of the
hadrons. Chiral symmetry breaking also
is only approximate, which will change the results for pion \cite{hmam}.
Also, old $SU(6)$ symmetry in spin and flavour space has been used for
the construction of proton and neutron states, where we know that this
symmetry is not sufficiently good. This shall have additional effects
not included here \cite{spm78,spm80a}.

Here $cos\chi(\vec k)$ through equations (2.33) and (2.34)
is really governed by the vacuum structure of QCD for chiral symmetry
breaking. The surprising feature is that this structure yields many
hadronic properties in the light quark sector in agreement with
experiments, leading to the conclusion that the low energy hadronic
properties are primarily
driven by the vacuum structure of quantum chromodynamics!

The results suggest that we should try to get $cos\chi(\vec k)$
through a variational procedure by minimising energy density parallel
to \cite{am91}. Such a programme would relate $R_{\pi}$ to the vacuum
structure of QCD where the value of $R_{\pi}$ will depend on $\alpha_s$
\cite{am91}. Implementation of the same will enable us to
recognise the coupling constant $\alpha_s$
for which it agrees with present phenomenological value, which will be
an additional conclusion from the present calculations.
\medskip

\newpage
\acknowledgements
The authors wish to thank N. Barik, A. R. Panda, H. Mishra, S.N. Nayak,
B. K. Parida and P. K.
Panda for many discussions. SPM acknowledges to the Department of Science
and Technology, Government of India for the project SP/S2/K-45/89.
AM would like to thank the Council of Scientific and Industrial
Research for a fellowship.

\def \ltg{ K.G. Wilson, Phys. Rev. D10, 2445 (1974); J.B. Kogut,
Rev. Mod. Phys. 51, 659 (1979); ibid 55, 775 (1983)}
\def \creutz { M. Creutz, Phys. Rev. Lett. 45, 313 (1980); ibid
Phys. Rev. D21, 2308 (1980)}
\def \svz {M.A. Shifman, A.I. Vainshtein and V.I. Zakharov,
Nucl. Phys. B147, 385, 448 and 519 (1979);
R.A. Bertlmann, Acta Physica Austriaca 53, 305 (1981)}
\def \spmbst {S.P. Misra, Phys. Rev. D35, 2607 (1987)}
\def \hmgrnv { H. Mishra, S.P. Misra and A. Mishra,
Int. J. Mod. Phys. A3, 2331 (1988)}
\def \snss {A. Mishra, H. Mishra, S.P. Misra
and S.N. Nayak, Phys. Lett 251B, 541 (1990)}
\def \amqcd { A. Mishra, H. Mishra, S.P. Misra and S.N. Nayak,
Pramana (J. of Phys.) 37, 59 (1991); A. Mishra, H. Mishra, S.P. Misra
and S.N. Nayak, Zeit. fur Phys. C, to appear}
\def \spmtlk {S.P. Misra, Talk on `Phase transitions in quantum field
theory' in the Symposium on Statistical Mechanics and Quantum field theory,
Calcutta, January, 1992, IP/BBSR/92-60}
\def \hmnj {H. Mishra and S.P. Misra, IP/BBSR/92-26}
\def \amcrl {A. Mishra, H. Mishra and S.P. Misra, IP/BBSR/92-52, to
appear in Z. Phys. C}
\def \higgs { S.P. Misra, in {\it Phenomenology in Standard Model and Beyond},
Proceedings of the Workshop on High Energy Physics Phenomenology, Bombay,
edited by D.P. Roy and P. Roy (World Scientific, Singapore, 1989), p.346;
A. Mishra, H. Mishra, S.P. Misra and S.N. Nayak, Phys. Rev. D44, 110 (1991)}
\def \nmtr {A. Mishra,
H. Mishra and S.P. Misra, Int. J. Mod. Phys. A5, 3391 (1990); H. Mishra,
 S.P. Misra, P.K. Panda and B.K. Parida, to appear in Int. J. Mod. Phys. E }
\def \dtrn {P.K. Panda, R. Sahu and S.P. Misra,
Phys. Rev C45, 2079 (1992)}
\def \qcd {G. K. Savvidy, Phys. Lett. 71B, 133 (1977);
S. G. Matinyan and G. K. Savvidy, Nucl. Phys. B134, 539 (1978); N. K. Nielsen
and P. Olesen, Nucl.  Phys. B144, 376 (1978); T. H. Hansson, K. Johnson,
C. Peterson Phys. Rev. D26, 2069 (1982)}
\def \cornwal {J.M. Cornwall, Phys. Rev. D26, 1453 (1982)}
\def \mndglv {J. E. Mandula and M. Ogilvie, Phys. Lett. 185B, 127 (1987)}
\def \schwinger {J. Schwinger, Phys. Rev. 125, 1043 (1962); ibid,
127, 324 (1962)}
\def \schutte {D. Schutte, Phys. Rev. D31, 810 (1985)}
\def \gft{ For gauge fields in general, see e.g. E.S. Abers and
B.W. Lee, Phys. Rep. 9C, 1 (1973)}
\def \gribov {N. Gribov, Nucl. Phys. B139, 1 (1978)}
\def \spmftm {S.P. Misra, Phys. Rev. D18, 1661 (1978); {\it ibid}
D18, 1673 (1978)}
\def \lopr {A. Le Youanc, L.  Oliver, S. Ono, O. Pene and J.C. Raynal,
Phys. Rev. Lett. 54, 506 (1985)}
\def \spphi {S.P. Misra and S. Panda, Pramana (J. Phys.) 27, 523 (1986);
S.P. Misra, {\it Proceedings of the Second Asia-Pacific Physics Conference},
edited by S. Chandrasekhar (World Scientfic, 1987) p.369}
\def \spmstr { S.P. Misra, Phys. Rev. D21, 1231 (1980)}
\def \spmjet {S.P. Misra, A.R. Panda and B.K. Parida, Phys. Rev Lett.
45,322 (1980); S.P. Misra and A.R. Panda, Phys. Rev. D21, 3094 (1980)}
\def \arpftm {L. Maharana, A. Nath and A.R. Panda, Mod. Phys. Lett. 7,
2275 (1992)}
\def \van {R. Van Royen and V.F. Weisskopf, Nuov. Cim. 51A, 617 (1965)}
\def \rchpi {S.R. Amendolia {\it et al}, Nucl. Phys. B277, 168 (1986)}
\def \chrl{ Y. Nambu, Phys. Rev. Lett. 4, 380 (1960);
 J.R. Finger and J.E. Mandula, Nucl.Phys.B199, 168 (1982);
A. Amer, A. Le Yaouanc, L. Oliver, O. Pene and
J.C. Raynal, Phys. Rev. Lett. 50, 87 (1983);
ibid, Phys. Rev. D28, 1530 (1983); S.L. Adler and A.C. Davis,
Nucl. Phys. B244, 469 (1984); R. Alkofer and P. A. Amundsen,
Nucl. Phys. B306, 305 (1988);
B. Haeri and M.B. Haeri, Phys. Rev. D43, 3732 (1991);
S. Li, R. S. Bhalerao and R. K. Bhaduri, Int. J. Mod. Phys.
A6, 501 (1991);
V. Bernard, Phys. Rev. D34, 1601 (1986)}
\def \spmijp { S.P. Misra, Ind. J. Phys. 61B, 287 (1987)}

\end{document}